\documentclass{Interspeech}

\interspeechcameraready

\title{LASPA: Language Agnostic Speaker Disentanglement with Prefix-Tuned
Cross-Attention}

\author[equalcontribution]{Aditya}{Srinivas Menon}
\author[equalcontribution]{Raj Prakash}{Gohil}
\author{Kumud}{Tripathi}
\author{Pankaj}{Wasnik}

\affiliation[nocounter]{}{Media Analysis Group}{Sony Research India}
\email{\{aditya.menon, raj.gohil, kumud.tripathi, pankaj.wasnik\}@sony.com}

\usepackage{comment}
\usepackage{subcaption}
\begin{document}

\maketitle

% the abstract here must exactly match the abstract entered into the paper submission system
\begin{abstract}
    
    % 1000 characters. ASCII characters only. No citations.
    % Speaker recognition models face significant challenges in multi-lingual settings due to the entanglement of linguistic information within speaker embeddings. The overlap between vocal traits such as accent, vocal anatomy, and language's phonetic structure complicates the separation of linguistic and speaker information. Effectively disentangling these components can significantly improve speaker recognition accuracy in multi-lingual contexts. To this end, we propose a novel disentanglement learning strategy that integrates joint learning through prefix-tuned cross-attention. This approach proves particularly effective in scenarios where speakers switch between multiple languages. Experimental results demonstrate the model's capability to generalize across monolingual and multi-lingual settings, including previously unseen languages. Notably, the proposed model significantly improves the equal error rate across multiple datasets, underscoring its ability to disentangle language information from the speaker embeddings. 
    Speaker recognition models face challenges in multi-lingual settings due to the entanglement of linguistic information within speaker embeddings. The overlap between vocal traits such as accent, vocal anatomy, and a language’s phonetic structure complicates separating linguistic and speaker information. Disentangling these components can significantly improve speaker recognition accuracy. To this end, we propose a novel disentanglement learning strategy that integrates joint learning through prefix-tuned cross-attention. This approach is particularly effective when speakers switch between languages. Experimental results show the model generalizes across monolingual and multi-lingual settings, including unseen languages. Notably, the proposed model improves the equal error rate across multiple datasets, highlighting its ability to separate language information from speaker embeddings and enhance recognition in diverse linguistic conditions.
\end{abstract}

\section{Introduction}
\begin{figure*}[!ht]
    \centering
    \includegraphics[width=1\linewidth, height=0.4\linewidth]{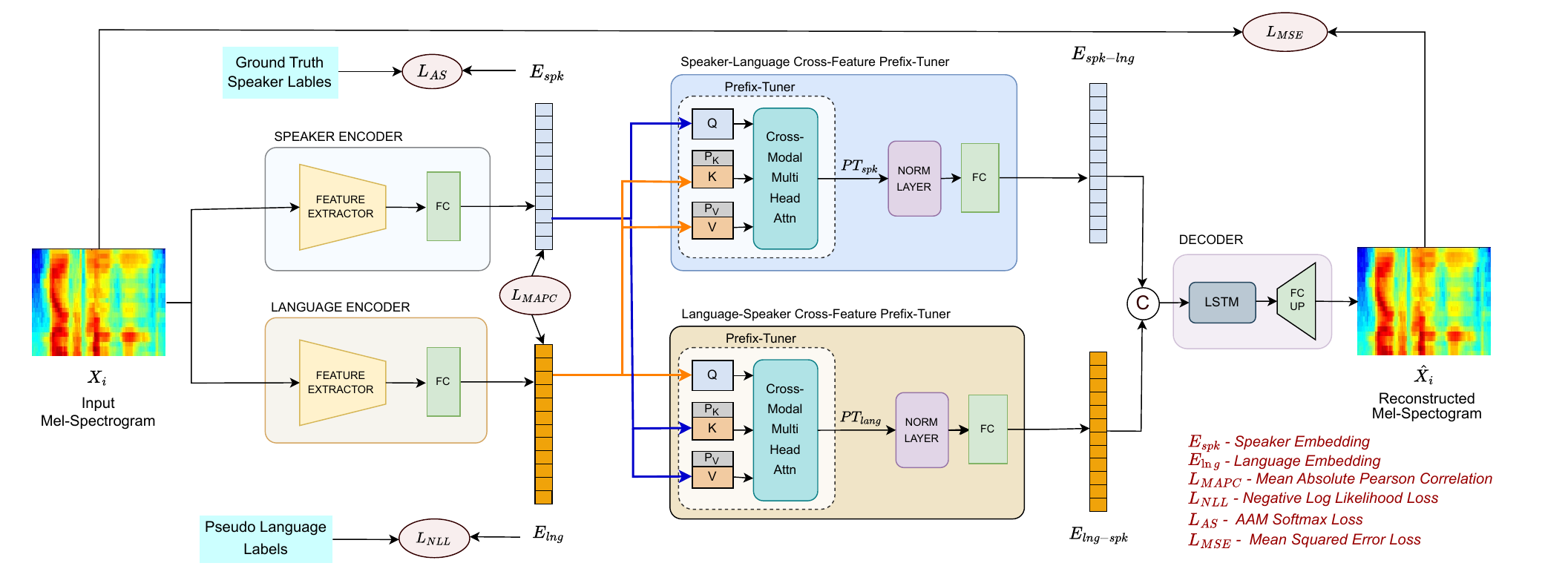} % Use \linewidth to span the full width
    \caption{Block diagram of the proposed architecture. Speaker and Language embeddings are extracted using the speaker and language encoder respectively and then after it is fused through prefix-tuners, and then passed to the decoder for mel-spectogram reconstruction.}
    \label{fig:architecture}
\end{figure*}
Speaker recognition in multi-lingual contexts presents significant challenges due to the complex interaction between language-specific characteristics, speaker identity, and acoustic variability. When speakers use different languages, their phonetic and prosodic patterns are influenced by the linguistic structure of each language and can vary considerably leading to these challenges. As a result, traditional speaker recognition systems often misinterpret these variations as different speakers, failing to recognize the same individual across languages. This issue arises due to the entanglement of language-dependent and speaker-dependent acoustic features within the speech signal ~\cite{nam23_interspeech, maiti2020generating}. Linguistically, each language imposes unique articulatory and prosodic requirements, which cause systematic variation in a speaker’s voice. Features such as phoneme structure, intonation, and co-articulation strategies differ significantly across languages, leading to distinct acoustic behavior. These differences, while linguistically driven, are not indicative of a change in speaker identity. However, speaker recognition models often fail to distinguish between language-induced variability and speaker-specific traits, reducing accuracy when identifying speakers across multiple languages.

Disentangled representation learning has been proposed as an effective approach to address this issue. The goal is to separate speaker identity-related features from language-dependent acoustic characteristics, which should remain invariant across languages. This method improves speaker recognition in multi-lingual settings by learning representations robust to linguistic variation, thereby mitigating the impact of language on recognition performance~\cite{nam23_interspeech}. Specifically, disentangled representation learning allows the model to isolate speaker-specific traits—such as vocal-tract characteristics, prosody, and voice timbre while suppressing language-dependent features like phonetic and prosodic variation. This separation helps the system focus on what remains consistent across languages, enabling more accurate cross-lingual speaker recognition.

% To overcome this, we present gradient reversal learning-free language-agnostic speaker embeddings learned using prefix-tuned cross-attention. Additionally, we proposed a multi-task learning strategy to learn the language disentangled speaker embeddings using prefix tuned cross attention while just using pseudo language labels. 
Traditionally, adversarial learning strategies based on the Gradient Reversal Layer (\textit{GRL}) have been proposed to disentangle speaker-specific information from language-specific acoustic features~\cite{ganin2015unsupervised, mun2022disentangled, yi2022disentangled, tong22_odyssey, kang2022augmentation}. However, \textit{GRL}-based approaches are prone to hyperparameters and can frequently lead to unstable training \cite{nam23_interspeech}. To address these challenges, previous work has proposed \textit{GRL}-independent approaches for effective disentangled representation learning~\cite{arjovsky2017towards, kang2020disentangled, wang2022disentangled}. Motivated by these approaches, we introduce {LASPA}, a multi-task learning strategy utilizing prefix-tuned cross attention, designed to achieve robust separation of speaker and language features while ensuring stable and efficient training. Our key contributions are summarized as follows:

\begin{enumerate}
    \item The paper proposes a joint learning-based approach to derive language-independent speaker embeddings for multi-lingual speaker recognition and speaker diarization.
    \item Speaker and language encoders, along with prefix-tuners are employed to fuse embeddings, ensuring accurate signal reconstruction by the decoder.
    % \item To enhance the learning of speaker embedding, this work uses multiple loss functions like \textit{MAPC}, \textit{AAM Softmax}, \textit{MSE}, and  \textit{NLL}.
    \item To enhance the learning of speaker embedding, this work uses multiple loss functions like Mean Absolute Pearson’s Correlation (\textit{MAPC})~\cite{kang2020disentangled}, Additive Angular Margin Softmax (\textit{AAM Softmax})~\cite{b19}, Mean Squared Error (\textit{MSE}), and Negative Log Likelihood (\textit{NLL}).
\end{enumerate}
\section{Previous Work}
Previous studies have shown that speaker representations are often entangled with factors such as emotions~\cite{williams2019disentangling, x-vectors}, accent~\cite{wang2021accent}, age~\cite{Raj_2019}, and environmental conditions like noise and reverberation~\cite{Campbell_628714}. More recently, research has validated that linguistic features embedded in speaker representations make them susceptible to language variability, leading to inaccurate recognition of the same speaker across different languages~\cite{nam23_interspeech, maiti2020generating}. While state-of-the-art (SOTA) models such as ResNet~\cite{b21, b23} and ECAPA (Emphasized Channel Attention, Propagation, and Aggregation)~\cite{b24} perform well in general speaker recognition tasks, they struggle in multi-lingual settings where a single speaker switches between languages, primarily due to the presence of linguistic information in the speaker representation.

Recently, there has been growing interest in speaker recognition tasks in multi-lingual scenarios. In this direction, \cite{nam23_interspeech} addressed the challenge of speaker recognition in bilingual scenarios by introducing VoxCeleb1-B, a large-scale evaluation set derived from VoxCeleb1, which is designed to test models in bilingual settings. They also propose a novel disentangled representation learning strategy that combines \textit{GRL} with \textit{MAPC} minimization. This approach shows significant improvements in performance for bilingual speaker recognition. Alongside the GRL, the latest SOTA ReDimNet \cite{yakovlev24_interspeech} uses the advantages of mapping 2D-1D convulation blocks to get more robust embeddings across different scenarios.

With the growing need for more efficient models, \cite{b29} introduced prefix-tuning as a lightweight alternative to traditional fine-tuning methods in large pre-trained models like GPT and BERT. Instead of updating all model parameters, prefix-tuning adds task-specific "prefix vectors" to the input sequence, keeping the original model weights frozen. Following its success in NLP, prefix-tuning has been extended to cross-modal applications.~\cite{ghadiya2024cross} claims that the prefix-tuned bottleneck attention helps in efficient multi-modal interaction between different modalities. It takes into account the importance of one modality to the other while performing fusion. These advances highlight the potential for prefix-tuning in multi-task learning.
\begin{table*}[ht]
    \centering
    \caption{EER (\%) and minDCF on cleaned versions of the VoxCeleb1 test sets, VoxCeleb1-B, VoxSRC 2021, VoxSRC 2020 validation set, and NISP-B. The two version of LASPA with and without prefix-tuner is compared with other state-of-the-art models}
    \begin{tabular}{lcccccc}
    \toprule
    \textbf{Model} & \multicolumn{2}{c}{\textbf{VoxSRC 2021 Val}} & \multicolumn{2}{c}{\textbf{VoxCeleb1-B}} & \multicolumn{2}{c}{\textbf{NISP-B}} \\
    \cmidrule(lr){2-3} \cmidrule(lr){4-5} \cmidrule(lr){6-7}
    & \textbf{EER ($\downarrow$)} & \textbf{minDCF ($\downarrow$)} & \textbf{EER ($\downarrow$)} & \textbf{minDCF ($\downarrow$)} & \textbf{EER ($\downarrow$)} & \textbf{minDCF ($\downarrow$)} \\
    \midrule
    ResNet-S(Baseline) & 9.22 & 0.500 & 9.69 & 0.618 & 22.33 & 0.993 \\
    GRL ResNet-S & 8.35 & 0.460 & 8.25 & 0.510 & -- & -- \\
    \textbf{LASPA ResNet-S} & \textbf{6.90} & \textbf{0.415} & \textbf{5.88} & \textbf{0.319} & \textbf{14.33} & \textbf{0.668} \\
    \midrule
    ResNet-L (Baseline) & 5.16 & 0.308 & 5.96 & 0.397 & 21.13 & 0.973 \\
    GRL ResNet-L & 4.22 & 0.246 & 3.69 & 0.254 & -- & -- \\
    \textbf{LASPA ResNet-L} & \textbf{3.12} & \textbf{0.212} & \textbf{2.38} & \textbf{0.102} & \textbf{14.75} & \textbf{0.558} \\
    \midrule
    ECAPA (Baseline) & 4.42 & 0.292 & 4.90 & 0.325 & 20.60 & 0.727 \\
    \textbf{LASPA ECAPA} & \textbf{3.01} & \textbf{0.195} & \textbf{2.07} & \textbf{0.086} & \textbf{11.95} & \textbf{0.544} \\
    \midrule
    ReDimNet (Baseline) & 2.15 & 0.086 & 1.66 & 0.083 & 11.90 & 0.607 \\
    \textbf{LASPA ReDimNet} & \textbf{2.10} & \textbf{0.079} & \textbf{1.62} & \textbf{0.053} & \textbf{10.22} & \textbf{0.233} \\
    \bottomrule
    \end{tabular}
    
    \vspace{0.5cm}
    
    \begin{tabular}{lcccccccc}
    \toprule
    \textbf{Model} & \multicolumn{2}{c}{\textbf{VoxCeleb1 cl.}} & \multicolumn{2}{c}{\textbf{VoxCeleb1-E cl.}} & \multicolumn{2}{c}{\textbf{VoxCeleb1-H cl.}} & \multicolumn{2}{c}{\textbf{VoxSRC 2020 Val}} \\
    \cmidrule(lr){2-3} \cmidrule(lr){4-5} \cmidrule(lr){6-7} \cmidrule(lr){8-9}
    & \textbf{EER ($\downarrow$)} & \textbf{minDCF ($\downarrow$)} & \textbf{EER ($\downarrow$)} & \textbf{minDCF ($\downarrow$)} & \textbf{EER ($\downarrow$)} & \textbf{minDCF ($\downarrow$)} & \textbf{EER ($\downarrow$)} & \textbf{minDCF ($\downarrow$)} \\
    \midrule
    ResNet-S (Baseline) & 2.27 & 0.167 & 2.43 & 0.175 & 4.74 & 0.299 & 6.91 & 0.390 \\
    GRL ResNet-S & 2.15 & 0.170 & 2.42 & 0.171 & 4.49 & 0.284 & 6.54 & 0.370 \\
    \textbf{LASPA ResNet-S} & \textbf{1.96} & \textbf{0.155} & \textbf{2.38} & \textbf{0.156} & \textbf{4.22} & \textbf{0.184} & \textbf{6.22} & \textbf{0.256} \\
    \midrule
    ResNet-L (Baseline) & 1.17 & 0.083 & 1.30 & 0.091 & 2.58 & 0.164 & 4.06 & 0.231 \\
    GRL ResNet-L & 0.99 & 0.079 & 1.25 & 0.088 & 2.42 & 0.154 & 3.91 & 0.220 \\
    \textbf{LASPA ResNet-L} & \textbf{0.95} & \textbf{0.063} & \textbf{1.05} & \textbf{0.086} & \textbf{2.19} & \textbf{0.119} & \textbf{3.42} & \textbf{0.212} \\
    \midrule
    ECAPA (Baseline) & 1.24 & 0.089 & 1.46 & 0.093 & 2.66 & 0.162 & 4.42 & 0.233 \\
    \textbf{LASPA ECAPA} & \textbf{0.89} & \textbf{0.045} & \textbf{1.00} & \textbf{0.076} & \textbf{2.01} & \textbf{0.095} & \textbf{3.21} & \textbf{0.192} \\
    \midrule
    ReDimNet (Baseline) & 0.37 & 0.030 & 0.53 & 0.051 & 1.00 & 0.097 & 1.66 & 0.128 \\
    \textbf{LASPA ReDimNet} & \textbf{0.31} & \textbf{0.026} & \textbf{0.49} & \textbf{0.047} & \textbf{0.86} & \textbf{0.087} & \textbf{1.42} & \textbf{0.095} \\
    \bottomrule
    \end{tabular}
    
    \label{tab:results-cross-lingual}
\end{table*}

\section{Model Architecture}
%Our proposed joint learning architecture to disentangle language-related information from speaker representation. 
The architectural overview of the proposed system is shown in Fig.~\ref{fig:architecture}. The architecture comprises a Speaker Encoder, a Language Encoder, two Prefix-Tuners, and a Decoder. The input waveform is initially resampled to a $16$ kHz sampling rate. Subsequently, mel-spectrogram is computed using a Hamming window with a $25$ ms window size and a $10$ ms stride. These mel-spectrograms $X_i$ serve as the input for both the Speaker Encoder and Language Encoder.

\subsection{\textbf{Speaker Encoder}}
\label{sec:adapter}
The proposed Speaker Encoder includes a feature extractor that transforms the mel-spectrogram $X_i$ into a $J$ dimensional vector. The output from the feature extractor is then passed through a feed-forward layer. The final output of this process is considered as the speaker embedding $E_{spk}$.
% \section{Introduction}
% \label{sec:introduction}

\subsection{\textbf{Language Encoder}}
Similarly, the proposed Language Encoder includes a feature extractor that transforms the mel-spectrogram $X_i$ into a $J$ dimensional vector. The final output of this process is considered as the language embedding $E_{lng}$.

\subsection{\textbf{Prefix-Tuner}}
To facilitate interaction between speaker and language features and extract relevant information, we employ two prefix-tuners~\cite{b29}. Speaker-Language cross-feature Prefix-Tuner $PT_{spk}$ and Language-Speaker cross-feature Prefix-Tuner $PT_{lang}$ for speaker and language information, respectively. 

% These prefix tuners integrate prior knowledge into the feature transformation process by blending learned representations with initialized parameters. For the speaker prefix tuner, the approach works by appending the keys $K$ and values $V$ derived from the language features $\it E_{lng}$ with prefixes $P_k$ and $P_v$, leading to the creation of prefix-tuned keys $K_p$ and values $V_p$. The prefix parameters $P_k$ and $P_v$ are initially set to zero matrices. These prefix-enhanced keys $K_p$ and values $V_p$, together with the query $Q$, represented by the speaker features $\it E_{spk}$, are then fed into the cross-feature multi-head attention module. 

% Similarly, for the language prefix tuner ($PT_{lang}$), the keys ($K$) and values ($V$) are derived from the speaker features ($E_{spk}$). These are appended with corresponding prefix terms ($P_k$ and $P_v$), resulting in the prefix-tuned keys ($K_p$) and values ($V_p$). The query $Q$ in this case, represented by the language features ($E_{lng}$), is then fed to the cross-feature multi-head attention module. This mechanism enables selective focus on the most relevant aspects of both speaker and language information, enhancing the interaction between these two features.

These prefix-tuners blend initialized parameters with learned representations to enhance feature interaction. In the Language-Speaker cross-feature Prefix-Tuner, the query $Q$ comes from language embedding $E_{lng}$, while the keys $K$ and values $V$ are derived from speaker embedding $E_{spk}$, appended with prefix terms $P_k$ and $P_v$. Conversely, in the Speaker-Language cross-feature Prefix-Tuner, the query comes from speaker features, and the keys and values are derived from language features $E_{lng}$ with the corresponding prefixes $P_k$ and $P_v$. Both tuners use these prefix-enhanced keys and values in cross-feature multi-head attention~\cite{NIPS2017_3f5ee243}, enabling focused interaction between speaker and language information.
The attention is obtained by comparing the query with the prefix-tuned keys and values. Mathematically, this is expressed in Equation \ref{eq1}:
\begin{align}
\label{eq1}
    F_{Att} = \text{Softmax}\left(\frac{Q \cdot K_p^T}{\sqrt{J_{K_p}}}\right) \cdot V_p
\end{align}
% \[
% F_{Att} = \text{Softmax}\left(\frac{Q \cdot K_p^T}{\sqrt{J_{K_p}}}\right) \cdot V_p
% \]
where \( J_{K_p} \) represents the dimensionality of the key vectors $K_p$. The resulting attention features $F_{Att}$ are passed to the prefix-tuners give the output as $E_{spk-lng}$ and $E_{lng-spk}$

%This adapter facilitates smooth integration between features while preserving feature-specific traits. 

% The design of the adapter module ensures efficient adjustment of input features, encouraging context-sensitive fusion across the modalities while maintaining their unique characteristics.

\subsection{Decoder}
The prefix-tuned speaker and the language embedding, $E_{spk-lng}$ and $E_{lng-spk}$ respectively, are concatenated and passed to the decoder that attempts to reconstruct the mel-spectrogram input $X_i$. The Decoder consists LSTM~\cite{lstm} followed by an Multi-layer perceptron(MLP) layer~\cite{popescu2009multilayer} to upscale and reconstruct the mel-spectrogram $\hat{X}_i$

\section{Training and Inference}
During training, the input waveform is converted into mel-spectrograms, denoted as $X_i$. In the Speaker Encoder, the mel-spectrogram passes through a feature extraction module and a fully connected layer. Details of the feature extractor can be found in Section~\ref{sec:baseline}. The output of the fully connected layer is the speaker embedding, $E_{spk}$. Similarly, in the Language Encoder, the mel-spectrogram passes through a feature extractor and fully connected layer to obtain the language embedding, $E_{lng}$.

For optimization, the \textit{AAM Softmax} loss function, $L_{AS}$, is applied to the speaker embedding, $E_{spk}$, using the ground truth speaker label to compute the speaker classification loss. Simultaneously, the \textit{NLL} loss function, $L_{NLL}$, is applied to the language embedding, $E_{lng}$, with pseudo-language labels to obtain the language classification loss. Furthermore, a mutual metric-based information loss, $L_{MAPC}$, is imposed between $E_{spk}$ and $E_{lng}$ to encourage disentanglement. Finally, we concatenate these embeddings and pass them through a decoder which consists of an LSTM module and a feed-forward network, to reconstruct the mel-spectrogram. A \textit{MSE} loss, $L_{MSE}$, is applied between the reconstructed mel-spectrogram, $\hat{X}i$, and the input mel-spectrogram. The proposed model is optimized using the final loss defined as:
\begin{align}
    L = L_{MSE} + L_{AS} + L_{MAPC} + L_{NLL}
\end{align}
% \[
% L = L_{MSE} + L_{AS} + L_{MAPC} + L_{NLL}
% \]
We empirically found that using equal weights led to optimal performance across evaluation metrics. During inference, we mainly use the Speaker Encoder, excluding the language encoder and cross-feature prefix-tuners. This ensures rapid embedding generation while maintaining high performance and language disentanglement. 

\label{sec:res}
This section provides detailed information about the implementation and datasets, followed by the evaluation and results obtained. The best results in the tables are highlighted in bold.
% Learning rate 1e-3, with optimizer as adam with weight decay of 2e-5. we use stepLR with gamma=0.95 and step-size =15
% RESNET-S=1.4M and RESNET-L=8M
\subsection{Implementation Details}
% This section outlines the implementation and parameter values used for the experiments. 
% Uniform parameter values were employed in all experiments, unless stated otherwise. 
% RESNET-S=1.4M and RESNET-L=8M
% We conducted our experiments using 4 NVIDIA A100 (40GB) GPUs. Learning rate 1e-3, with optimizer as Adam\cite{zhang2018improved} with weight decay of 2e-5. we use stepLR with gamma=0.95 and step-size =15 
We opted for two versions of ResNet i.e., ResNet-S with $1.4$ million and ResNet-L with $8$ million parameters, ECAPA and ReDimNet with $15$ million parameters - as our backbone feature extractor in LASPA as it shows significant improvement across various datasets~\cite{yakovlev24_interspeech}.  Our experiments were carried out using $8$ NVIDIA A100 GPUs ($40$ GB each) server. The models were trained using an Adam optimizer with a weight decay of $2e-5$, and the learning rate was set at $1e-3$. Each batch consisted of processing $400$ utterances at a time and each configuration is trained for $1200$ epochs. %Each utterance
% frame is During training, clips were uniformly sampled from videos, while during evaluation, a sliding window with half overlap was utilized to extract the samples.
% % Other module weights were initialized randomly. 
% The model was trained using ASAM~\cite{asam} with AdamW~\cite{adamw} as the base optimizer with $\rho=2$. 
% % The AdamW optimizer utilized a weight decay of $0.05$ and an initial learning rate of $0.001$. 
% A cosine annealing learning rate scheduler was employed with a warm-up of 3 epochs and an initial learning rate of 1e-5 during the warm-up. For the SoftIC loss, a feature dimension of 128 and a memory bank size of 256 were used, with a loss weight (i.e., $\lambda_{SIC}$) of $0.001$. Mixup~\cite{mixup} augmentation was applied to the training samples with a mixing factor of $\alpha=0.1$. Additionally, random horizontal flips, random Gaussian blur, and random adjustments to brightness, contrast, and saturation-based augmentations were used in the training process. 
% The proposed method is implemented in PyTorch and trained on 4 A100 GPUs with a batch size of 2 on each GPU.
% % Training was performed with a minimum batch size of 2 on each GPU. 
% At a time, 128 frames were processed, constituting a single sample (clip). During training, clips were uniformly sampled from videos, while during evaluation, a sliding window with half overlap was utilized to extract the samples.

\subsection{Datasets}
For training, we use the VoxCeleb2~\cite{chung18b_interspeech} dev set, a benchmark dataset commonly used for speaker recognition tasks consisting of $5,994$ speakers. To obtain language-agnostic speaker representations, we leverage language pseudo-labels from the VoxCeleb2 dataset using a Spoken Language Recognition (\textit{SLR}) model trained on the VoxLingua107~\cite{b28} dataset. For testing, we employ three original monolingual test sets from VoxCeleb1~\cite{nagrani2017voxceleb}, VoxSRC2020, and the multi-lingual VoxSRC2021 validation set~\cite{b23}, along with VoxCeleb1-B~\cite{nam23_interspeech}. The model is tested on the multi-lingual NISP-B dataset to further evaluate the method on out-of-distribution data. 
% \begin{table}[h]
%    \centering
%     \caption{\textsc{Overview of Test Sets Featuring Cross-Lingual Pairing for Evaluation.}}

%     \begin{tabular}{lcccccc}
%         \toprule
%         \textbf{Dataset} & \textbf{Cross-lingual} \\
%         \midrule
%         VoxSRC 2021 Val & \cmark \\
%         VoxSRC 2020 Val & \xmark \\
%         VoxCeleb1 cl. & \xmark \\
%         VoxCeleb1-E cl. & \xmark \\ 
%         VoxCeleb1-H cl. & \xmark \\ 
%         VoxCeleb1-B & \cmark \\
%         NISP-B & \cmark \\
%         \bottomrule
%     \end{tabular}

%     \label{tab:cross-lingual}
% \end{table}
We create the NISP-B dataset from NISP~\cite{b31}, a multi-lingual speaker profiling dataset that features speech recordings in English, Hindi, Kannada, Malayalam, Tamil, and Telugu. This dataset includes intra-speaker cross-lingual trials and inter-speaker monolingual trials, with balanced representation across all languages.

% Evaluating the model's capability for language disentanglement in languages not included in the VoxCeleb2 dataset provides a more robust and comprehensive assessment of its generalization. By extending the testing to unseen languages, we can better ascertain whether the model has truly learned to produce language-agnostic speaker embeddings rather than overfitting onto the specific languages in the training data. Table~\ref{tab:cross-lingual} presents a detailed overview of the diverse test sets utilized in our evaluation, highlighting whether each set incorporates cross-lingual pairs for testing purposes.
% This evaluation provides a more comprehensive understanding of the model's ability to generalize and disentangle language across both monolingual and multilingual environments.
% Table~\ref{tab:cross-lingual} outlines the diverse test sets used, including cross-lingual pairs, to assess the model’s performance across monolingual and multilingual settings.

\section{Experimental Analysis}
\label{sec:baseline}

\subsection{Baselines}

We use ResNet-(S~\cite{b21} and L~\cite{b23}), ECAPA~\cite{b24} and ReDimNet~\cite{yakovlev24_interspeech} as speaker encoder architectures which are trained from scratch. For the Language Encoder we use ECAPA architecture and is initialized from the model pre-trained on VoxLingua107. We evaluate using Equal Error Rate (\textit{EER}) and minimum decision cost function (\textit{minDCF}).

\subsection{Results}

Our experiments demonstrate the effectiveness of the proposed architecture in disentangling language information from speaker embeddings, leading to improved performance across various test sets. As shown in Table~\ref{tab:results-cross-lingual}, our approach LASPA, consistently outperforms baseline models on VoxCeleb1-B, VoxSRC 2021 validation, and NISP-B cross-lingual test sets. Specifically, ReDimNet, ResNet-L, and ECAPA architectures with our disentanglement method achieve the lowest \textit{EER} and \textit{minDCF} values, with LASPA ReDimNet reducing the \textit{EER} to $2.10$ \% on VoxSRC 2021 validation and $1.62$ \% on VoxCeleb1-B. Moreover, LASPA ReDimNet achieves notable performance gains with an \textit{EER} of $10.22$ \% on NISP-B, a challenging multi-lingual dataset.

In Table~\ref{tab:results-cross-lingual}, we observe similar trends, with our models consistently achieving lower \textit{EER} and \textit{minDCF} values on cleaned versions of monolingual VoxCeleb1, VoxCeleb1-E, VoxCeleb1-H, and VoxSRC 2020 validation sets. These results confirm that the disentanglement strategy improves the robustness of speaker embeddings, enabling language-agnostic performance across multi-lingual datasets.

\begin{table}[t]
       \centering
       \caption{Spoken Language Recognition (\%) on VoxCeleb1-B, Cosine Similarity scores on VoxCeleb1 cl. and Diarization Error Rate (\%) on DISPLACE dataset across models.}
        \setlength{\tabcolsep}{1.5pt}
        \begin{tabular}{lcccc}
               \toprule

               \textbf{Model} & \textbf{SLR Accuracy ($\downarrow$)} & \textbf{Similarity ($\uparrow$)} & \textbf{DER ($\downarrow$)} \\
               \midrule
               DISPLACE \cite{baghel23_interspeech} & -- & -- & 32.5\\
               ResNet-S    & 88.4 & 0.348 & 30.4 \\ 
               ResNet-L & 89.2 & 0.391 & 29.2 \\
               ECAPA  & 91.1 & 0.531 & 28.7 \\
               ReDimNet & 81.2 & 0.612 & 28.5 \\
               \textbf{LASPA} & \textbf{78.5} & \textbf{0.641} & \textbf{28.2} \\
               \bottomrule
           \end{tabular}
           
            \label{tab:more_results}
\end{table}

The \textit{SLR} metric refers to the ability of a system to correctly identify the language being spoken from an audio signal. Table~\ref{tab:more_results} demonstrates the observed decrease in \textit{SLR} accuracy from speakers in the VoxCeleb1-B dataset, indicating less language-specific information in $E_{spk}$. Table~\ref{tab:more_results} also shows the cosine similarity between audios of the same speaker across languages. Higher scores indicate robust and consistent speaker embeddings for individual speakers, confirming that LASPA enhances speaker distinction. Column 3 of the table shows the Diarization Error Rate (\textit{DER}) on the DISPLACE dataset ~\cite{baghel23_interspeech, 10.1007/978-3-031-48312-7_39}, showing that our approach outperforms the DISPLACE baseline.

\subsection{Ablation Study}

To further analyze the contribution of each component in LASPA, we conducted an ablation study, focusing on the impact of the language encoder and prefix-tuning.  We compared the full LASPA model with two reduced configurations: one without prefix-tuning (but with the language encoder and decoder), and a "Speaker-only" model (excluding language-related components). We performed this experiment using ResNet-S as the speaker encoder model.

Table~\ref{tab:ablation} presents the Equal Error Rate (EER) results on the VoxCeleb1-C cleaned and VoxCeleb1-B datasets for these configurations.
\begin{table}[t]
    \centering
    \caption{Ablation Study Results reporting EER (\%) on monolingual VoxCeleb1 cl. and bilingual VoxCeleb1-B dataset}
    \label{tab:ablation}
        \setlength{\tabcolsep}{4pt}
    \begin{tabular}{lcc}
        \toprule
        \textbf{Configuration} & \textbf{VoxCeleb1 cl.} ($\downarrow$) & \textbf{VoxCeleb1-B} ($\downarrow$)\\
        \midrule
        LASPA (Full Model) & \textbf{1.96} & \textbf{5.88} \\
        No Prefix-Tuning & 2.14 & 6.44 \\
        Speaker-only & 2.27 & 7.00\\
        \bottomrule
    \end{tabular}
\end{table}
As shown in Table~\ref{tab:ablation}, the LASPA model achieves the lowest EER on VoxCeleb1 cl. and VoxCeleb1-B. Both models that integrate language information significantly outperform the "Speaker-only" baseline. Additionally, prefix-tuners account for just $1.16$ \% of the model's total parameters, making them a highly efficient way to enhance performance.

\section{Conclusion}
%We propose a method for generating language-agnostic speaker embeddings using prefix-tuned cross-attention, addressing the challenge of disentangling language from speaker embeddings in multi-lingual speaker recognition. Our approach avoids Gradient Reversal Learning (GRL), leading to more stable training. Evaluations on multiple datasets, including out-of-distribution languages, show its robustness and strong generalization. The method improves both speaker recognition and diarization. Future work could extend it to more languages, other speech tasks, and disentangling factors like accent and style for further gains.

In this work, we introduced LASPA, a novel approach for language-agnostic speaker disentanglement using prefix-tuned cross-attention. Our method effectively separates speaker identity from linguistic information, addressing key challenges in multi-lingual speaker recognition. By integrating prefix-tuning, we achieve efficient adaptation while using only a small fraction of the model’s total parameters. Experimental results demonstrate that LASPA consistently improves speaker recognition performance across both monolingual and multi-lingual scenarios, including unseen languages. The reduction in equal error rate (EER) across multiple datasets validates the effectiveness of our approach in mitigating the influence of language on speaker embeddings. %Furthermore, our method is particularly beneficial in code-switching scenarios, where speakers alternate between languages. 
Given its efficiency and strong generalization capability, LASPA offers a promising direction for robust speaker recognition in diverse linguistic environments. Future work will explore further refinements and applications of prefix-tuning in speaker verification and related tasks.
\pagebreak

\bibliographystyle{IEEEtran}
\bibliography{mybib}

\end{document}